\input{aipcheck}

\documentclass[
    ,final            
  ]
  {aipproc}

\layoutstyle{8x11single}

\begin{document}

\title{A remarkable representation of the Clifford group}

\author{Ingemar Bengtsson}{
  address={Fysikum, Stockholm University, 106 91 Stockholm, Sweden}
}


\begin{abstract}
The finite Heisenberg group knows when the dimension of Hilbert 
space is a square number. Remarkably, it then admits a representation 
such that the entire Clifford group---the automorphism group of the 
Heisenberg group---is represented by monomial phase-permutation 
matrices. This has a beneficial influence 
on the amount of calculation that must be done to find Symmetric 
Informationally Complete POVMs. I make some comments on the equations 
obeyed by the absolute values of the components of the SIC vectors, 
and on the fact that the representation partly suggests a preferred 
tensor product structure. 

%
%
\end{abstract}

\maketitle


\section{The Heisenberg group}

The name "Heisenberg group" for the group we are interested in became 
widely used some 50 years after Heisenberg did the path breaking work 
that Arkady Plotnitsky describes elsewhere in this volume---and which 
is indeed closely related to this group. The very foundations of 
quantum mechanics are closely related to the Heisenberg group. 

Actually there are many Heisenberg groups. They all admit a defining 
representation in terms of upper triangular matrices of the form 

\begin{equation} g(\phi, x,p) = \left( \begin{array}{ccc} 
1 & x & \phi \\ 0 & 1 & p \\ 0 & 0 & 1 \end{array} \right) 
\ . \end{equation}

\noindent Such matrices form a nilpotent group as soon as the 
entries $x$, $p$, and $\phi$ can be added and multiplied together 
in such a way that they form a group under addition. In other 
words, the entries must belong to a ring, but the ring can be 
chosen freely. If the entries are real numbers we obtain a three 
dimensional Lie group, whose Lie algebra includes the commutator 
between position and momentum---figuring so prominently in 
Heisenberg's discoveries. If the entries belong to the ring of integers 
modulo $N$ we obtain a finite Heisenberg group, denoted 
$H(N)$, which has an essentially unique unitary irreducible 
representation in a Hilbert space of $N$ dimensions. Other options 
are available, such as choosing the entries to belong to a finite 
field. This leads to finite groups represented in $p^k$-dimensional 
Hilbert spaces, where $p$ is a prime number. Unless $N$ is a prime 
number the latter groups differ from $H(N)$. 

I have no intention to describe this zoo of groups in detail here. 
Suffice it to say that it played a major role in two of the 
most brilliant episodes of human thinking, nineteenth century 
geometry \cite{Edd} and twentieth century physics---and that I believe 
they will play a major role in this century too. The focus 
in my talk was on the role they play in the problem of finding 
Symmetric Informationally Complete POVMs, or SICs for short. 
In the laboratory SICs correspond to a special kind of (doable 
\cite{Steinberg}) measurements, of interest for the foundations of 
quantum mechanics \cite{Chris}. My personal motivation for working 
on the problem will be revealed at the end. 

A SIC is a collection of $N^2$ unit vectors in a Hilbert 
space of dimension $N$, such that they form a resolution of the 
identity, i.e. a POVM, and such that 

\begin{equation} |\langle \psi_I|\psi_J\rangle |^2 = \frac{1}{N+1}
(1 + N\delta_{IJ}) \ , \hspace{8mm} 1 \leq I,J \leq N^2 \ .   
\label{SIC} \end{equation}  

\noindent Finding such collections of vectors is surprisingly hard. 
Zauner's conjecture states that they do exist for any $N$, that they 
can be chosen such that they form an orbit of the Heisenberg 
group $H(N)$, and such that every vector in the collection is 
left invariant under a special order three element of the 
automorphism group of $H(N)$ \cite{Zauner}. This automorphism group 
is what we call the Clifford group here. The evidence for Zauner's 
conjecture comes from explicit constructions, not from any structural 
understanding. Scott and Grassl give a wonderful summary of the 
status of this conjecture in the concluding section of their recent 
paper \cite{Scott}. 

It is known that $H(N)$ is the only group that can serve the purpose 
if $N$ is a prime not equal to 3 \cite{Huangjun}. It is also known 
that the analogous problem concerning equiangular lines in real Hilbert 
spaces does not have a solution in general \cite{equi}. 

There is a problem with a similar flavour, which is that of finding a 
complete set of Mutually Unbiased orthonormal Bases, or MUB. Here it is 
believed that a solution exists only if the dimension equals a prime 
number or a power of a prime number. The existing solutions \cite{Wootters} 
are again closely related to the Heisenberg groups, but then one uses 
the version based on finite fields, which is not the same as $H(N)$ unless 
the dimension $N$ is a prime number. It is odd that the SIC and MUB 
existence problems differ in this way. They also differ in that---at least 
from one point of view---the latter is partly understood: 
it is known that the existence of a complete set of MUB is equivalent 
to the existence of a unitary operator basis forming a flower with 
$N+1$ petals, in the sense that the $N^2$ operators in the basis can 
be divided into $N+1$ petals of mutually commuting operators, such that 
two different petals have only the unit element in common \cite{Boykin}. 
If we restrict ourselves to unitary operator bases whose elements form 
a group, this means that we must look for groups with a sufficient 
number of maximal abelian subgroups. It is known that no group can do 
better than the Heisenberg groups in this regard \cite{Asch}. (Why one 
should restrict oneself to unitary operator bases of group type is not 
clear at all. But this is beside the point for the moment.)  

\section{Square dimensions}

It would seem as if $H(N)$, the Heisenberg group over the ring of integers 
modulo $N$, has a kind of "universal" structure that is insensitive to the 
choice of $N$, except for some special features that emerge when 
$N$ is prime. However, closer inspection reveals that there is something 
very special about square dimensions too. 

To see this, recall that the usual presentation of $H(N)$ is in terms 
of the root of unity 

\begin{equation} \omega = e^{\frac{2\pi i}{N}} \ , \end{equation}

\noindent and generators $X$ and $Z$ subject to the relations 

\begin{equation} ZX = \omega XZ \ , \hspace{8mm} X^N = Z^N = {\bf 1} \ . 
\end{equation}

\noindent When presented like this the Heisenberg group admits a unique 
unitary representation---up to unitary equivalence \cite{Weyl}. The usual 
choice then is to use the representation where the cyclic subgroup 
generated by $Z$ is diagonal. But suppose that the dimension $N = n^2$ 
is a square number. Then it is readily seen that 

\begin{equation} Z^nX^n = X^nZ^n \ . \end{equation}

\noindent This gives rise to a preferred maximal abelian subgroup 
all of whose $N$ elements are of order $n$. What we \cite{pp} refer to as the 
phase permutation basis is the basis in which this special 
subgroup is diagonal. Its basis vectors are labelled $|r,s\rangle $, 
where $r$ and $s$ are integers modulo $n$, and we define the phase 
factor 

\begin{equation} q = e^{\frac{2\pi i}{n}} = \omega^n \ . \end{equation}

\noindent With the phase conventions 
we used the representation of the group generators is then 

\begin{equation} X|r,s\rangle = \left\{ \begin{array}{ll} |r,s+1\rangle 
& \mbox{if} \ s+1 \neq 0 \ \mbox{mod} \ n \\
\ \\
q^r|r, 0\rangle & \mbox{if} \ s + 1 = 0 \ \mbox{mod} \ n \end{array} \right. 
\label{WHrep} \end{equation} 

\begin{equation} Z|r,s\rangle = \omega^s |r-1,s\rangle \ . \nonumber 
\end{equation}

\noindent This leads to 

\begin{equation} X^n|r,s\rangle = q^r|r,s\rangle \ , 
\hspace{8mm} Z^n|r,s\rangle = q^s|r,s\rangle \ . \label{8} \end{equation}

\noindent All matrices here are monomial unitaries, also known as  phase-permutation 
matrices, meaning that exactly one element in each column is non-zero, 
and equal to a phase factor, and similarly for the rows. Actually this 
much is true for the usual representation too. The reason why we refer to 
the new basis as the phase-permutation basis \cite{pp} is that 
its use has in its train that the entire Clifford 
group is given by phase-permutation matrices ---and this is not at all true 
for the usual representation of the Heisenberg group. Recall that the Clifford 
group consists of all unitary operators $U$ such that $UX^iZ^jU^\dagger$ 
is again an element of the Heisenberg group---and that the Clifford group 
plays a very important role when one tries to understand the SIC problem, 
which is our main concern for the moment.

To show that the Clifford group is represented by phase-permutation 
matrices it is enough to observe that, being an automorphism 
group, it must permute the maximal abelian subgroups---known as stabiliser 
groups in quantum information theory---among themselves. But it must 
also preserve the order of all group elements. When the dimension equals 
$n^2$ there exists a unique stabiliser group consisting solely of elements 
of order $n$. This is the stabiliser group that determines the new basis. 
The Clifford group can only permute its elements among themselves,  
and the remarkable property of the basis follows immediately. 
 
There is an interesting connection to the theory of theta functions here; 
in particular the `new' representation is used by Mumford in his presentation 
of the theory of theta functions from the point of view of the Heisenberg 
group \cite{Mumford}. The Heisenberg group over the real numbers can be 
represented by generators $S_x$ and $T_y$ acting on the space of analytic 
functions through 

\begin{equation} (S_xf)(z) = f(z+x) \ , \hspace{8mm} (T_yf)(z) = 
e^{\pi i y^2\tau + 2\pi i yz}f(z+\tau y) \ , \end{equation}

\noindent where $\tau$ is a complex number with positive imaginary 
part. If we restrict ourselves to the lattice of integer valued 
$x$ and $y$ there is a unique function invariant under these 
transformations, namely the theta function 

\begin{equation} \theta (z,\tau) = \sum_{k = - \infty}^\infty 
e^{\pi i k^2\tau + 2\pi i kz} \ . \end{equation}

\noindent Using a net with somewhat finer meshes we catch an invariant 
$n^2$-dimensional vector space of functions. This vector space admits a 
natural basis consisting of theta functions of rational characteristic. 
Thus we define a subgroup of the real Heisenberg group by insisting that 
$na$ and $nb$ are integers. Let $\theta_{0,0}(z) = \theta (z)$. As a basis 
for the vector space we take the functions 

\begin{equation} \theta_{a,b}(z) = S_bT_a\theta (z) \ , \hspace{8mm} na, nb \in 
{\bf Z} \ . \end{equation}
 
\noindent The phase-permutation representation is recovered from the observations 
that

\begin{equation} S_{c} \theta_{a,b} = \theta_{a,b+c} \ , \hspace{8mm} 
T_c\theta_{a,b} = e^{-2\pi ibc}\theta_{a+c,b} \ , \hspace{5mm} na,nb,nc \in 
{\bf Z}  \end{equation}

\noindent and 

\begin{equation} \theta_{a+x,b+y} = e^{2\pi iay}\theta_{a,b} \ , \hspace{8mm} 
x,y,na,nb \in {\bf Z} \ . \end{equation}

\noindent The reader may be familiar with the case $n = 1/2$, discussed in 
Whittaker and Watson \cite{WW}; in a notation closer to Jacobi's we have $(\theta_{0,0}, \theta_{0,\frac{1}{2}}, \theta_{\frac{1}{2},0}, \theta_{\frac{1}{2},
\frac{1}{2}}) = (\theta_3, \theta_4, \theta_2, \theta_1)$. Via elliptic curves 
the theta functions play a prominent role in the geometry of complex projective 
space, so the connection may prove helpful for the SIC problem, or in some other 
way---although at the moment this is a vague suggestion only. 

In another direction, note the temptation to write $|r,s\rangle = |r\rangle 
\otimes |s\rangle$. Eq. (\ref{8}) then says that the special subgroup we 
are basing the representation on is represented by local operators, that is 
by operators of the form $U_1\otimes U_2$. Moreover the generator $Z$ is a 
local operator---as indeed it was in the representation proposed by Weyl 
\cite{Weyl}, where $Z$ is diagonal. If then the basis vectors are separable 
one finds 

\begin{equation} Z = \mbox{diag}(1,\omega, \dots ,\omega^{N-1}) = 
\mbox{diag}(1,q,\dots , q^n)\otimes \mbox{diag}(1,\omega , \dots, 
\omega^{n-1}) \ . \end{equation}

\noindent The unitary that connects the two representations is a local operator 
with respect to this tensor product structure, so this property is preserved. 
We can conclude, for instance, that the $n^4$ vectors in a given SIC must fall in 
$n$ multiplets with the same degree of entanglement in each. (Curiously, 
even more is true for some $N = 4$ SICs: all the 16 vectors have the 
same concurrence \cite{Berge}. I do not understand 
why.) However, note that our two requirements---that $|r,s\rangle$ forms a 
phase-permutation basis and that $|r,s\rangle = |r\rangle 
\otimes |s\rangle$---contains some ambiguities. The Main Property of the 
phase-permutation basis is that 
it leads to a representation of the automorphism group of the Heisenberg 
group by phase-permutation matrices, and this property is kept if the basis 
is changed by an arbitrary unitary phase-permutation matrix. In general 
this will disrupt the tensor product structure. Still the 
phase-permutation basis partly determines the tensor product structure 
of the Hilbert space, and this may turn out to be interesting---but again 
this is a somewhat vague idea at the moment. 

\section{The SIC equations for the absolute values}

I advertised that the phase-permutation basis simplifies the SIC 
problem to some extent. The main reason is that it is well adapted to 
exploit Zauner's conjecture, according to which any SIC vector is left 
invariant by a special element of the Clifford group. This does help, 
although not as much as one could hope for. The details can be found 
in the paper that we wrote \cite{pp}---and since the paper appeared very recently 
I cannot add much to the story here. 

What I can add are a few comments that were left out of our paper. In a 
previous V\"axj\"o meeting Marcus Appleby discussed a subset of the SIC 
equations (\ref{SIC}), which concern only the absolute values of the 
components of the SIC vectors \cite{Appleby}. He assumed that the 
SIC is an orbit of the Heisenberg group. In this case the 
SIC is completely determined by a single fiducial vector $|\psi_0\rangle$ 
such that  

\begin{equation} |\langle \psi_0|X^iZ^j|\psi_0\rangle |^2 = \left\{ 
\begin{array}{lll} 1 & \ & i = j = 0 \\ \ \\ \frac{1}{N+1} & & 
\mbox{otherwise} \ . \end{array} \right. \end{equation}

\noindent The equations involve quartic polynomials in $N$ variables, 
and most known solutions have been derived using Gr\"obner bases and 
dedicated computer programs such as MAGMA \cite{Scott}. Still it is interesting 
to manipulate them a bit further, in order to see what is involved. Let 
the components of the fiducial vector be given by 

\begin{equation} z_a = \sqrt{p_a}e^{i\mu_a} \ . \end{equation}

\noindent In the usual representation it is known that the SIC equations 
are equivalent to the equations \cite{Mahdad, ADF} 

\begin{eqnarray} \sum_{a=0}^{N-1}p_a^2 = \frac{2}{N+1} \ , \hspace{24mm} 
\label{cond1} \\ \nonumber  \\ 
\sum_{a=0}^{N-1}p_ap_{a+x} = \frac{1}{N+1} \ , \hspace{8mm} x \in \{1, 2, \dots 
{N-1}\} \ , \label{cond2} \\ \nonumber \\  
\sum_{a=0}^{N-1}\bar{z}_a\bar{z}_{a+k-i}z_{a+k}z_{a-i} = 0 \ , \hspace{8mm} i,k \neq 0 
\ . \hspace{2mm} \label{cond3} \end{eqnarray} 

\noindent Interestingly there are $N$ equations that do not involve the phases 
at all. Moreover these equations have an interesting geometrical interpretation, 
arising when we project all vectors onto the simplex spanned by the basis 
vectors in a cross section of the convex body of density matrices. The equations 
imply that the image of the $N^2$ SIC vectors is a regular simplex of a 
characteristic size, inscribed in the larger simplex \cite{Appleby}. 
Unfortunately the geometric interpretation also suggests that the 
absolute values $p_i$ cannot be determined by these equations alone, since 
the orientation of the image simplex is not determined. 

If we examine eqs. (\ref{cond1}-\ref{cond2}) we see that they they remain 
the same if the sign of the integer $x$ is changed. Hence there are at most 
$k+1$ equations independent equations, regardless of whether $N = 2k$ or 
$N = 2k+1$. We also see that they imply 

\begin{equation} \left( \sum_{a=0}^{N-1}p_a \right)^2 = 1 \ . 
\label{19} \end{equation}

\noindent So normalisation is included---as in fact it was from the start, 
before the SIC equations were brought to this form. If $N = 2k$ we can also 
derive that 

\begin{equation} \left( \sum_{r=0}^{k-1}p_{2r} - \sum_{r=0}^{k-1}p_{2r+1} 
\right)^2 = \frac{1}{N+1} \ , \label{20} \end{equation}

\noindent which again is easily imposed, and means that the equations can be 
linearised if $N = 2$. However, I cannot find any further relation that 
strikes the eye and is easy to handle.

Very similar equations can be deduced in the phase permutation basis, where 
the absolute values are naturally denoted by $\sqrt{p_{rs}}$. Recall that the 
dimension is $N = n^2$. From the equations for a Heisenberg covariant SIC 
one deduces that \cite{pp}

\begin{equation} \sum_{r,s=0}^{n-1}p_{rs}p_{rs} = \frac{2}{N+1} 
\ \ , \label{moduli1} \end{equation}

\begin{equation} \sum_{r,s=0}^{n-1}p_{rs}p_{r+x,s+y} = \frac{1}{N+1} 
\ . \label{moduli2} \end{equation}

\noindent Here $x,y$ are integers modulo $n$, not both zero. In itself this 
is a slight improvement on the previous result; the number of independent 
equations goes up to $k+2$ if $N = 2k$. This means that the absolute 
values are determined by these equations alone if $N = 4$. 

Let $N = 4$. Then there are four equations for the absolute values, all 
of which can be brought to a form analogous to eqs. (\ref{19}-\ref{20}):

\begin{equation} \left\{ \begin{array}{l} p_{00}^2 + p_{01}^2 + p_{10}^2 
+ p_{11}^2 = \frac{2}{5} \\ \\ 2p_{00}p_{01} + 2p_{10}p_{11} = \frac{1}{5} \\ \\  
2p_{00}p_{10} + 2p_{01}p_{11} = \frac{1}{5} \\ \\ 2p_{00}p_{11} + 2p_{01}p_{10} = \frac{1}{5} \end{array} 
\right. \hspace{5mm} \Leftrightarrow \hspace{5mm} 
\left\{ \begin{array}{l} (p_{00} + p_{01} + p_{10} + p_{11})^2 
= 1 \\ \\ (p_{00}+p_{01}-p_{10}-p_{11})^2 = \frac{1}{5} \\ \\  
(p_{00}+p_{10}-p_{01}-p_{11})^2 = \frac{1}{5} \\ \\ 
(p_{00}+p_{11}-p_{01}-p_{10})^2 = \frac{1}{5} \end{array} 
\right.\ . \end{equation}

\noindent After taking the square roots we obtain a linear system of equations. 
Since the components appear symmetrically in the equations we may assume without 
loss of generality that $p_{00} \geq p_{01} \geq p_{10} \geq p_{11} \geq 0$. This determines three of the signs. If the sign of the fourth square root is 
positive one finds that 

\begin{equation} p_{01} = p_{10} = p_{11} = \frac{5 - \sqrt{5}}{20} \ . \end{equation}

\noindent The negative sign is inconsistent with $p_{rs} \geq 0$. Hence the moduli 
are completely determined, and solving for the phases afterwards is not too hard. 
Notice that the calculation is not only straightforward, it also gives every 
Heisenberg covariant SIC in four dimensions, with no input from numerical 
searches.

The phase-permutation basis comes into its own when we impose Zauner's 
conjecture. By some further permutations and rephasings of the basis vectors one 
can block diagonalise a Zauner unitary, so that it consists of three by three 
blocks of the form 

\begin{equation} \left( \begin{array}{ccc} 0 & 0 & 1 \\ 1 & 0 & 0 \\ 0 & 1 & 0 
\end{array} \right) \ , \end{equation}

\noindent together with some diagonal elements. Each such block contributes 
three eigenvalues $(1, e^{2\pi i/3}, e^{4\pi i/3})$ to the spectrum. Given that 
the spectrum of a Zauner unitary is known \cite{Zauner} one can deduce that 
there will be 3 diagonal elements
if $N = 3l$, and only one if $N = 3l + 1$. Regardless of whether the block 
diagonalisation is carried through, one concludes that an 
invariant vector will have $l+1$ independent components only, 
where $N = n^2 = 3l$ or $3l+1$. The resulting equations 
(\ref{moduli1}-\ref{moduli2}) for the absolute values are then easy to 
manage, at least for $N = 9, 16, 25$ which we did explicitly. But it turns 
out that the number of independent equations is too small to determine 
all the absolute values---and solving the equations that involve the 
phases is not easy at all \cite{pp}. In fact it has never been done, in 
any basis, for $N = 16, 25$.\footnote{This is what I said in my talk. 
Eventually things worked out better than I expected---the final version of 
our paper does contain the solution for $N = 16$ \cite{pp}.}
  
\section{Conclusion}

The phase-permutation basis partly suggests a preferred tensor product 
structure ${\cal H}^n\otimes {\cal H}^n$ of the Hilbert space ${\cal H}^N$. 
The commuting operators $X^n$ and $Z^n$ are local operators. This is 
interesting in itself. It also has the wonderful consequence that the entire 
Clifford group is represented by monomial unitary matrices. This 
simplifies the SIC existence problem, but more ideas are needed to 
trivialise it in dimensions $N > 4$. Perhaps some can be found.
 
All of this reminds me of Fermat's Last Theorem. The statement of Fermat's 
theorem is very simple. In itself it appears to be a curiousity only, but 
in the course of proving it mathematicians were led to develop deep 
theories that changed the face of their subject. 
I like to think that the SIC existence problem has a similar status within 
quantum mechanics.  
 
\begin{theacknowledgments}
I thank Andrei Khrennikov for once again arranging a successful conference, 
and Hulya and Marcus Appleby, Steve Brierley, Markus Grassl, 
David Gross, and Jan-\AA ke 
Larsson for allowing me to use some joint results for this contribution. 

\end{theacknowledgments}



\bibliographystyle{aipproc}   

\bibliography{sample}

\begin{thebibliography}{99}

\bibitem{Edd} I. Bengtsson, {\it From SICs and MUBs to Eddington}, J. Phys. Conf. 
Ser., to appear.

\bibitem{Steinberg} Z. E. D. Medendorp, F. A. Torres-Ruiz, L. K. Shalm, 
G. N. M. Tabia, C. A. Fuchs, and A. M. Steinberg, {\it Experimental characterization 
of qutrits using SIC-POVMs}, Phys. Rev. {\bf A83} (2011) 051801R.

\bibitem{Chris} C. A. Fuchs, {\it QBism, the Perimeter of quantum Bayesianism}, 
arXiv:1003.5209.

\bibitem{Zauner} G. Zauner: {\it Quantendesigns. Grundz\"uge einer 
nichtkommutativen Designtheorie}, PhD thesis, Univ. Wien 1999. Available 
in English translation as {\it Quantum Designs: Foundations of a noncommutative 
Design Theory}, Int. J. Quant. Inf. {\bf 9} (2011) 445.

\bibitem{Scott} A. J. Scott and M. Grassl, {\it Symmetric informationally complete 
positive-operator-valued measures}, J. Math. Phys. {\bf 51} (2010) 042203.

\bibitem{Huangjun} H. Zhu, {\it SIC-POVMs and Clifford groups in 
prime dimensions}, J. Phys. {\bf A43} (2010) 305305. 

\bibitem{equi} P. W. H. Lemmens and J. J. Seidel, {\it Equiangular lines}, J. Algebra {\bf 24} (1973) 494.

\bibitem{Wootters} W. K. Wootters and B. D. Fields, {\it Optimal state-determination 
by mutually unbiased measurements}, Ann. Phys. {\bf 191} (1989) 363. 

\bibitem{Boykin} S. Bandyopadhyay, P. O. Boykin, V. Roychowdhury, and F. Vatan, 
{\it A new proof for the existence of mutually unbiased bases}, Algorithmica {\bf 34} 
(2002) 512.

\bibitem{Asch} M. Aschbacher, A. M. Childs, and P. Wocjan, {\it The limitations of nice mutually unbiased bases}, J. Algebr. Comb. {\bf 25} (2007) 111. 

\bibitem{Weyl} H. Weyl: {\it Theory of Groups and Quantum Mechanics}, 
Dutton, New York 1932.

\bibitem{pp} D. M. Appleby, I. Bengtsson, S. Brierley, M. Grassl, D. Gross, 
and J.-\AA . Larsson, 
{\it The monomial representations of the Clifford Group}, Quantum Info. Comp, 
{\bf 12} (2012), to appear. 

\bibitem{Mumford} D. Mumford: {\it Tata Lectures on Theta I}, 
Birkh\"auser, Boston 1983.

\bibitem{WW} E. T. Whittaker and G. N. Watson: {\it A Course of Modern Analysis}, 
4th ed., Cambridge U.P. 1927.

\bibitem{Berge} H. Zhu, Y. S. Teo, and B.-G. Englert, {\it Structure of two-qubit Symmetric 
Informationally Complete POVMs}, Phys. Rev. {\bf A81} (2010) 052339.

\bibitem{Appleby} D. M. Appleby, {\it SIC-POVMs and MUBs: Geometrical relationships in 
prime dimensions}, in L. Accardi et al (eds.): Proc of the V\"axj\"o Conference 
on Foundations of Probability and Physics - 5, AIP Conf. Proc. 1101, New York 
2009.

\bibitem{Mahdad} M. Khatirinejad, {\it On Weyl-Heisenberg orbits of equiangular lines}, J. Algebr. Comb. {\bf 28} (2008) 333.

\bibitem{ADF} D. M. Appleby, H. B. Dang, and C. A. Fuchs, {\it Symmetric Informationally-Complete quantum states as analogues to orthonormal bases and 
Minimum Uncertainty States}, eprint arXiv:0707.2071.



\end{thebibliography}

\IfFileExists{\jobname.bbl}{}
 {\typeout{}
  \typeout{******************************************}
  \typeout{** Please run "bibtex \jobname" to optain}
  \typeout{** the bibliography and then re-run LaTeX}
  \typeout{** twice to fix the references!}
  \typeout{******************************************}
  \typeout{}
 }

\end{document}